\documentclass[twocolumn,epjc3]{svjour3} 

\journalname{Eur. Phys. J. A}
\usepackage{amssymb,latexsym}
\usepackage[utf8]{inputenc}
\usepackage{amsmath,amssymb,mathrsfs}
\usepackage{graphicx}
\usepackage[usenames]{color}
\usepackage{cite}
\usepackage{caption}

\usepackage{hyperref}
\hypersetup{colorlinks=true,linkcolor=blue,citecolor=blue,menucolor=black,urlcolor=blue,filecolor=blue}

\smartqed 
\RequirePackage{mathptmx} 
\RequirePackage[numbers,sort&compress]{natbib}

\newcommand{\dszero}{D_{s0}^*}
\newcommand{\dsone}{D_{s1}}
\newcommand{\bszero}{B^{*}_{s0}}
\newcommand{\bsone}{B_{s1}}
\newcommand{\pieta}{\pi^0{\rm -}\eta}

\begin{document} 

\title{Update on strong and radiative decays of the $\dszero(2317)$ and $\dsone(2460)$ and their bottom cousins
}

\author{Hai-Long Fu\thanksref{e1,add1,add2}, 
Harald W. Grie{\ss}hammer\thanksref{e2,add3},
Feng-Kun Guo\thanksref{e3,add1,add2}, 
Christoph~Hanhart\thanksref{e4,add4}, 
Ulf-G.~Mei{\ss}ner\thanksref{e5,add5,add4,add6}}

\thankstext{e1}{fuhailong@itp.ac.cn}
\thankstext{e2}{hgrie@gwu.edu}
\thankstext{e3}{fkguo@itp.ac.cn}
\thankstext{e4}{c.hanhart@fz-juelich.de}
\thankstext{e5}{meissner@hiskp.uni-bonn.de}

\institute{
  CAS Key Laboratory of Theoretical Physics, Institute of Theoretical Physics, Chinese Academy of Sciences, Beijing, 100190, China\label{add1}
  \and
  School of Physical Sciences, University of Chinese Academy of Sciences, Beijing, 100049, China\label{add2}
  \and 
  Institute for Nuclear Studies, Department of Physics, The George Washington University, Washington DC 20052, USA\label{add3}
  \and 
  Institute for Advanced Simulation, Institut f\"{u}r Kernphysik and J\"ulich Center for Hadron Physics, Forschungszentrum J\"{u}lich, D-52425 J\"{u}lich, Germany\label{add4}
  \and 
  Helmholtz-Institut f\"ur Strahlen- und Kernphysik and Bethe Center for Theoretical Physics, Universit\"at Bonn, D-53115 Bonn, Germany\label{add5}
  \and 
  Tbilisi State University, 0186 Tbilisi, Georgia\label{add6}
  }

\date{}

\maketitle

\begin{abstract}
  The isospin breaking and radiative decay widths of the positive-parity 
  charm-strange mesons, $D^{*}_{s0}$ and $D_{s1}$, and their predicted
  bottom-strange counterparts, $B^{*}_{s0}$ and $B_{s1}$, as hadronic molecules are revisited. This is necessary, since
  the $B^{*}_{s0}$ and $B_{s1}$ masses used in Eur. Phys. J. A \textbf{50} (2014) 149~\cite{Cleven:2014oka} were too small, in conflict with the heavy quark flavour symmetry. Furthermore, not all isospin breaking contributions were considered. We here present a method to restore
  heavy quark flavour symmetry, 
  correcting the masses of  $B^{*}_{s0}$ and $B_{s1}$, and include the complete isospin breaking contributions
  up to next-to-leading order. With this we
   provide updated  hadronic  decay widths for all of $D^{*}_{s0}$, $D_{s1}$, $B^{*}_{s0}$ and $B_{s1}$.
  Results for the partial widths of the radiative deays of $D_{s0}^*(2317)$ and $D_{s1}(2460)$ are also renewed in light of the much more precisely measured $D^{*+}$ width.
  We find that
   $B_s\pi^0$ and $B_s\gamma$ are the preferred channels for searching for $B_{s0}^*$ and $B_{s1}$, respectively.
\end{abstract}

\thispagestyle{empty}

The lightest positive-parity charm mesons $\dszero$~\cite{Aubert:2003fg} and $\dsone$~\cite{Besson:2003cp} are among the first observed heavy-flavour mesons with properties beyond quark model expectations. They are prominent candidates of hadronic molecules~\cite{Guo:2017jvc}, with the dominant components residing in the isoscalar $DK$ and $D^*K$ channels, respectively.
In particular, within the hadronic mo{\-}lecular picture, the approximate equality of the mass differences $M_{D^*}-M_D$ and $M_{\dsone}-M_{\dszero}$ is a natural consequence of heavy quark spin symmetry.
Yet, this might also be explained in the chiral doublet model of Refs.~\cite{Nowak:2003ra,Bardeen:2003kt}.
However, the partial decay widths of these states play a unique role in discriminating various models. In particular, the isospin breaking hadronic decays $\dszero\to D_{s}^{}\pi^{0}$ and $\dsone\to  D_{s}^{*}\pi^{0}$ are expected to be much larger in the molecular approach than in other models,
since in addition to the more conventional $\pi^0$-$\eta$ mixing mechanism,
also the mass differences of charged and
neutral constituents contribute via
loops. Moreover, these contributions are large only 
for molecular states and
enhanced by the nearby threshold cusps, a mechanism
well known from the $a_0$-$f_0$ mixing~\cite{Achasov:1979xc,Hanhart:2007bd}.

Heavy quark flavour symmetry (HQFS) allows one to estimate
the masses of the bottom partners of the $\dszero$ and $\dsone$, independent
 of the model assumptions for their internal structure.
 One finds
\begin{align}
  M_{\bszero} =&\, \bar M_c + \Delta_{b-c} + \left(M_{\dszero} - \bar M_c\right) \frac{m_c}{m_b} \simeq 5.71~\text{GeV}, \notag\\
  M_{\bsone} =&\, \bar M_c + \Delta_{b-c} + \left(M_{\dsone} - \bar M_c\right) \frac{m_c}{m_b} \simeq 5.76~\text{GeV},
  \label{eq:hqfs}
\end{align}
where $\bar M_c = (M_{\dszero} + 3 M_{\dsone})/4 = 2.42$~GeV is the spin-averaged mass of the charmed mesons, $\Delta_{b-c}$ is the difference between the bottom and charm quark mass scales,  which may be estimated by $\bar M_{B_s} - \bar M_{D_s}\simeq 3.33$~GeV, with $\bar M_{B_s}=5.40$~GeV and $\bar M_{D_s}=2.08$~GeV the spin-averaged masses of the ground state pseudoscalar and vector $Q\bar s$ mesons.
More precise results with controlled uncertainties in the hadronic molecular model can be found in Refs.~\cite{Albaladejo:2016lbb,Du:2017zvv} and are consistent with these simple predictions; see also Refs.~\cite{Guo:2006fu,Guo:2006rp}.
The predictions of Ref.~\cite{Kolomeitsev:2003ac} for these masses are
5761~MeV and 5807~MeV, respectively, about 50~MeV higher.
On the other hand,
the lattice results in Ref.~\cite{Lang:2015hza} are $M_{\bszero}^\text{lat}=(5711\pm13\pm19)$~MeV and  $M_{\bsone}^\text{lat}=(5750\pm17\pm19)$~MeV, also in nice agreement with the HQFS predictions~\eqref{eq:hqfs}. 

So far, neither of these two bottom-strange mesons has been observed, but they are currently searched for at the LHCb experiment. 
It is therefore important and timely to have reliable predictions of both radiative and isospin breaking hadronic decay widths of these mesons.
The predictions in the hadronic molecular model have been made in Refs.~\cite{Cleven:2014oka} (see also Refs.~\cite{Faessler:2007gv,Gamermann:2007bm,Faessler:2007us,Lutz:2007sk,Faessler:2008vc,Guo:2008gp,Xiao:2016hoa,Guo:2018kno} for predictions of some of the states and Refs.~\cite{Fajfer:2015zma,Fajfer:2016xkk} for calculations in chiral perturbation theory up to one-loop order). However, the $\bszero$ and $\bsone$ masses used there, $(5625\pm45)$~MeV and $(5671\pm45)$~MeV, respectively, are too low to be consistent with the HQFS expectations derived from Eq.~\eqref{eq:hqfs}.
Consequently,  the partial decay widths computed therein need to be corrected.

The low masses for the bottom strange scalar states
in Ref.~\cite{Cleven:2014oka} can be traced to
a heavy-quark mass independent
subtraction constant used to regularize the two-point scalar loop integral. However, 
a proper ma{\-}tching procedure unavoidably 
requires a heavy-quark mass dependence of this
quantity. To see this observe that the two-point loop integral, which includes the right-hand cut, is ultraviolet divergent and can be regularized in different ways. One option is to use dimensional regularization with a subtraction constant $a(\mu)$, where the regularization scale $\mu$ may be taken to be 1~GeV; variations of this value can be absorbed into the related
variation of $a(\mu)$. Alternatively,  the ultraviolet divergence can be regularized by a hard cutoff, $q_\text{max}$, on the magnitude of the loop three-momentum. We denote the resulting two-point scalar loop functions by $G_\text{sub}(s, a(\mu))$ and $G_\text{CO}(s, q_\text{max})$, respectively, with $s$ the c.m. energy squared.

In the previous analysis~\cite{Liu:2012zya}, 
where the low-energy constants were determined, the method
with a subtraction constant was used. 
However, the dependence of the subtraction constant on the heavy quark mass, $m_Q$,  is not clear a priori. In order to take this into account, it was proposed in Ref.~\cite{Guo:2006fu} to 
match $G_\text{sub}$ and $G_\text{CO}$ at the relevant charmed meson threshold to determine $q_\text{max}$ from a given $a(\mu)$, and then use the same $q_\text{max}$ in the bottom sector at a relevant bottom threshold to determine the value of $a(\mu)$ there.
Since the value of $q_\text{max}\sim 0.8$~GeV is much smaller than the charmed meson mass, HQFS should be well preserved in this way.\footnote{The procedure was also described in the previous study~\cite{Cleven:2014oka}; however, it was unfortunately not implemented in the code, and thus the masses of the $B_{s0}^*$ and $B_{s1}$ obtained and used therein were too small.}

To be specific, the expressions for the loop functions $G_\text{sub}(M_\text{thr}^2, a(\mu))$ and $G_\text{CO}(M_\text{thr}^2, q_\text{max})$  at the $HK$ threshold, $M_\text{thr} = M_H + M_K$, are given by
\begin{align}
&\, G_\text{sub}(M_\text{thr}^2, a(\mu)) = \frac{1}{16\pi^2}\left[a(\mu)+\frac{1}{M_\text{thr}} \sum_{i=H,K} M_{i}\log \frac{{M_{i}}^{2}}{\mu^2} \right], \notag\\
&\, G_\text{CO}(M_\text{thr}^2, q_\text{max}) \notag\\ =&\, \frac{1}{16\pi^2M_\text{thr}}\sum_{i=H,K} M_{i}\log \frac{{M_{i}}^{2}}{\left(\sqrt{M_{i}^{2}+q_\text{max}^{2}}+q_\text{max}\right)^2}.
\end{align}
The matching requires 
\begin{equation}
  G_\text{sub}(M_{\rm thr}^2, a(\mu)) = G_\text{CO}(M_{\rm thr}^2, q_\text{max}) \ ,
  \label{Dmatch}
\end{equation}
which leads to the heavy-meson-mass ($M_H$) dependence of $a(\mu)$ as
\begin{equation}
a(\mu)=\frac{1}{M_\text{thr}}\sum_{i=H,K} M_{i}\log \frac{{\mu}^{2}}{\left(\sqrt{M_i^{2}+q_\text{max}^{2}}+q_\text{max}\right)^2} .
\end{equation}
For the mass dependence of $a(\mu)$, see also Ref.~\cite{Oller:2000fj}.

In the charm sector we obtain for
$M_{\rm thr}=M_{D_s}+M_K$ and
 $a_D(1~\text{GeV}) = -1.87^{+0.05}_{-0.04}$~\cite{Liu:2012zya} 
 the three-momentum cutoff
\begin{equation}
  q_\text{max} = 745^{+35}_{-37} \ \rm{MeV} \ .
\end{equation}
With the same cutoff value, imposing the
matching condition of Eq.~(\ref{Dmatch})
with $M_{\rm thr}=M_{B_s}+M_K$
one finds
\begin{equation}
  a_B(1~\text{GeV}) = -3.41\pm0.02 \ . 
  \label{eq:ab}
\end{equation}
With this matching procedure, and considering the heavy-quark mass scaling of the low-energy constants (LECs)~\cite{Cleven:2010aw,Albaladejo:2016lbb}:
\begin{equation}
  \{h^{\prime B}_{i}\}\sim\{h^{\prime D}_{i}\}\frac{m_{B}}{m_{D}},
  \label{eq:scale}
\end{equation}
for $h_i' \in \{h_{24},h_{35}, h_4', h_5'\}$ (for their definitions, we refer to Ref.~\cite{Liu:2012zya}),
the $B^{*}_{s0}$ and $B_{s1}$ masses can be obtained in line with HQFS~\cite{Albaladejo:2016lbb,Du:2017zvv} as listed in Table~\ref{tab:masses}.

\begin{table}[tb]
  \begin{center}
  \caption{\label{tab:masses}Masses of the lowest positive-parity heavy-strange mesons~\cite{Albaladejo:2016lbb,Du:2017zvv} and their effective couplings, $g_1$ and $g_2$, respectively, to the isoscalar $D^{(*)}K$ and $D^{(*)}_s\eta$ ($\bar B^{(*)}K$ and $\bar B^{(*)}_s\eta$) channels predicted using the parameters in Ref.~\cite{Liu:2012zya} after the proper scaling of the subtraction constant.
  The mass of the $\dszero(2317)$ is fixed in obtaining the parameters. 
  The masses and couplings are given in units of MeV and GeV, respectively.
  }
  \begin{tabular}{lccccc}
  \hline\noalign{\smallskip}
  Meson &  Mass & $g_{1}$  & $g_{2}$ \\\noalign{\smallskip}
  \hline
          $D_{s0}^{*}$&  $2318$ (fixed) & $9.4\pm0.3$& $7.4\pm0.1$\\\noalign{\smallskip}
          $D_{s1}$ & $2458^{+15}_{-17}$ & $10.1^{+0.8}_{-0.9}$& $7.9\pm0.3$\\\noalign{\smallskip}
          $B_{s0}^{*}$ & $5722\pm14$ & $22.9^{+1.3}_{-1.5}$& $18.8^{+0.4}_{-0.5}$\\\noalign{\smallskip}
          $B_{s1}$ & $5774\pm13$ & $22.5^{+1.3}_{-1.5}$& $18.7\pm0.5$\\\noalign{\smallskip}
          \hline
  \end{tabular}
\end{center}
\end{table}

With the masses and the effective couplings, which are computed from the residues of the coupled-channel $T$-matrix, we can update the results in Ref.~\cite{Cleven:2014oka} of the partial decay widths of the $D_{s0}^*$ and $D_{s1}$ mesons by recalculating the Feynman diagrams therein in the hadronic molecular picture using the updated values of the involved meson masses and branching fractions\footnote{In particular, the $D^{*+}$ width is much more precisely known~\cite{Zyla:2020zbs}. The updated parameter values are: for the axial coupling $g_{\pi}=0.566\pm0.006$; for those related to the magnetic coupling, $1/\beta=335\ \rm MeV$, $m_{c}=1304 \ \rm MeV$, and $m_{b}=4660\ \rm MeV$. See Ref.~\cite{Cleven:2014oka} for the definitions of the parameters.} and correct the results for the $B_{s0}^*$ and $B_{s1}$.

\begin{figure}[h]
    \centering
    \includegraphics[width=\linewidth]{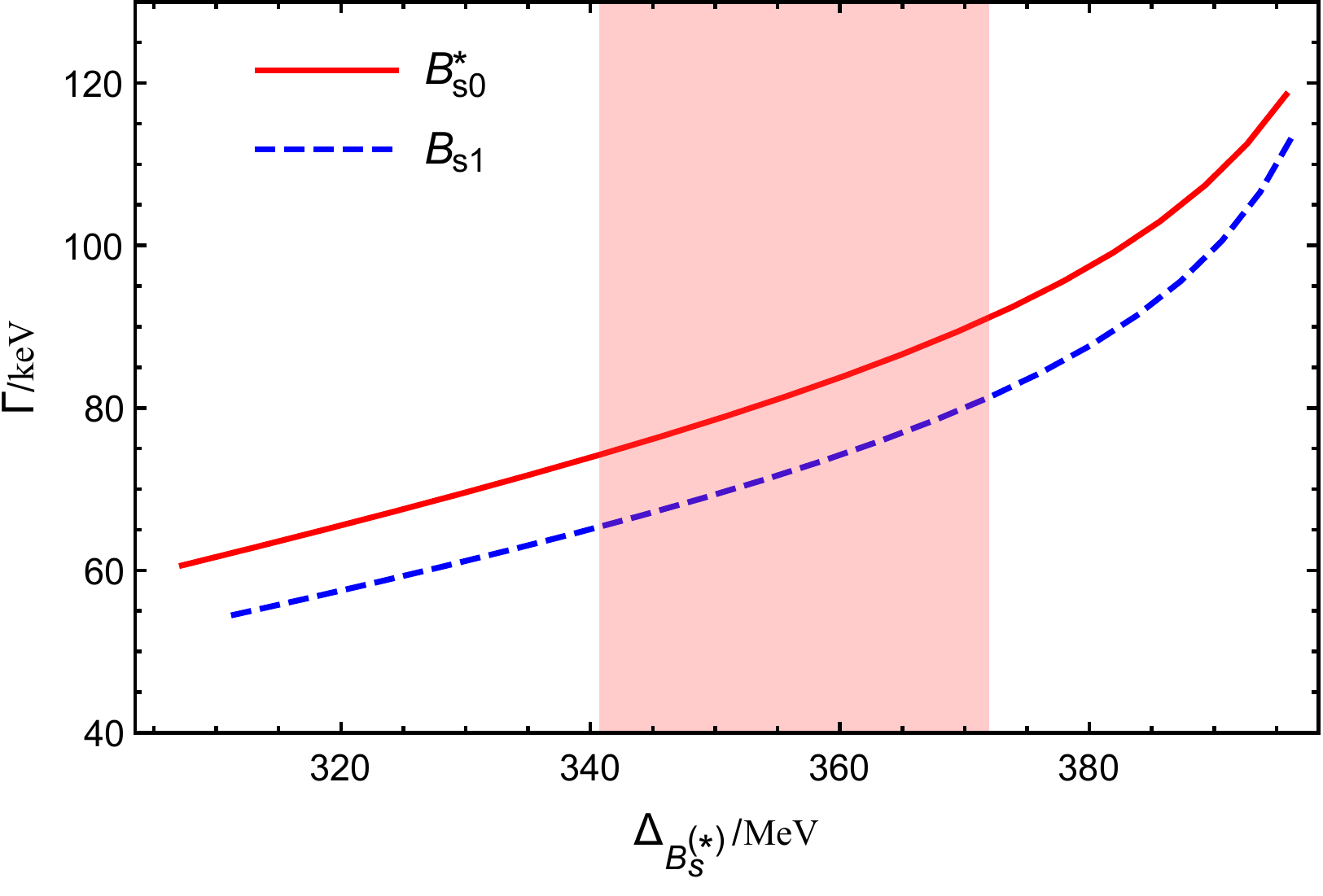}
    \caption{Dependence of the hadronic decay widths of the $B_{s0}^*$ and $B_{s1}$ on the mass differences $\Delta_{B_s}=M_{B_{s0}^*} - M_{B_s}$ and $\Delta_{B_{s}^*}=M_{B_{s1}} - M_{B_s^*}$, for the subtraction constant $a_B(1~{\rm GeV})$ in the region from $-3.31$ to $-3.51$, covering the range in Eq.~\eqref{eq:ab} (shaded in the plot). 
    }
    \label{fig:Delta}
\end{figure}

The results for the isospin breaking hadronic partial decay widths are shown in Table~\ref{tab:hadronicSU3f},\footnote{Since the widths are sensitive the masses of the involved particles, in order to get the proper hadronic decay width of the $D_{s1}(2460)$, the matching procedure described above should also be done at the $D^*K$ threshold to determine the value of $a(\mu)$ in this case. And the scaling in Eq.~\eqref{eq:scale} with $m_B$ replaced by $m_{D^*}$ is also necessary. Otherwise, the width of the $D_{s1}(2460)$ would be much smaller.} and those for the radiative partial decay widths are given in Table~\ref{tab:radiative}. Mainly because of the change of phase spaces, the results for the bottom mesons presented here, except for the $B_{s1}\to B_{s0}^*\gamma$ transition, are much larger than those in Ref.~\cite{Cleven:2014oka} that used too small $B_{s0}^*$ and $B_{s1}$ masses, and supersede the results therein.
In Fig.~\ref{fig:Delta}, we show the dependence of the hadronic decay widths of the $B_{s0}^*$ and $B_{s1}$ on the mass differences $\Delta_{B_s} = M_{B_{s0}^*} - M_{B_s}$ and $\Delta_{B_s^*} = M_{B_{s1}} - M_{B_s^*}$, respectively. 
One clearly sees the quick increase of the results as the mass differences get larger. 
From comparing the hadronic and radiative decay widths, one sees that the preferred decay modes for the $B_{s0}^*$ and $B_{s1}$ are $B_{s}\pi^0$,  the $B_s\gamma $, respectively. 

\begin{table}[tb]
  \centering
   \caption{Hadronic partial decay widths of $\dszero$, $\dsone$, $\bszero$ and $\bsone$ as hadronic molecules (in units of $\rm keV$) computed with the diagrams shown in Fig.~\ref{fig:Diagram}. The second and third columns correspond to the contributions from loops (Fig.~\ref{fig:Diagram} (a) and (b)), and the $\pi^0$-$\eta$ mixing (Fig.~\ref{fig:Diagram} (c)), respectively. Here the central values are obtained using the central values of the parameters $h_i$'s and $a(\mu)$ given in Table~VIII in Ref.~\cite{Liu:2012zya}, and the uncertainties are propagated from those of the parameters.}
  \begin{tabular}{lccc}
  \hline\noalign{\smallskip}    Decay channel & Loops & $\pi^{0}$-$\eta$ mixing & Full result                             \\\noalign{\smallskip}
  \hline 
  $D_{s0}^{*}\rightarrow D_{s}\pi^{0}$ &$50\pm3$ &$20\pm2$ &$132\pm7$ \\\noalign{\smallskip}
  $D_{s1}\rightarrow D_{s}^{*}\pi^{0}$ &$37\pm7$ &$20\pm3$ & $111\pm15$ \\\noalign{\smallskip}
  $B^{*}_{s0}\rightarrow B_{s}\pi^{0}$ &$15\pm2$  &$22\pm3$ &$75\pm6$\\\noalign{\smallskip}
  $B_{s1}\rightarrow B_{s}^{*}\pi^{0}$ &$16\pm2$ &$23\pm3$ &$76\pm7$ \\\noalign{\smallskip}
  \hline
  \end{tabular}
  \label{tab:hadronicSU3f}
\end{table}

\begin{table}[tb]
  \centering
  \caption{Radiative partial decay widths of $\dszero$, $\dsone$, $\bszero$ and $\bsone$ as hadronic molecules (in units of $\rm keV$). Here EC, MM and CT represent the contributions from only the electric coupling, the magnetic moment, and the contact term, respectively; ``$-$'' means that there is no such  contribution; ``?'' means that the corresponding contact term is unknown and the full result cannot be obtained without additional input. See Ref.~\cite{Cleven:2014oka} for details. Remark that we have recalculated the value of coupling of the contact term ($\kappa$ in Ref.~\cite{Cleven:2014oka}) and now its value is too small compared with the uncertainty.
  Therefore we did not list the uncertainty from the contact term contribution. }
  \begin{tabular}{lcccc}
  \hline\noalign{\smallskip}    Decay channel & EC& MM& CT & Full result \\\noalign{\smallskip}
  \hline
        $D^{*}_{s0}\rightarrow D^{*}_{s}\gamma$ &$3.5\pm0.3$ & $0.06\pm0.02$ &0.04 &$3.7\pm0.3$ \\\noalign{\smallskip}
        $D_{s1}\rightarrow D_{s}\gamma$ & $13\pm1$ &$6.5\pm0.6$ &0.1 &$42\pm4$ \\\noalign{\smallskip}
        $D_{s1}\rightarrow D^{*}_{s}\gamma$& $12\pm2$ &$0.8\pm0.1$ &0.1 & $13\pm2$ \\\noalign{\smallskip}
        $D_{s1}\rightarrow D^{*}_{s0}\gamma$&$-$& $3.0\pm0.6$ & ? & ? \\\noalign{\smallskip}
        $B^{*}_{s0}\rightarrow B^{*}_{s}\gamma$ & $58\pm8$ & $2.1\pm0.3$ & 0.02 & $59\pm8$ \\\noalign{\smallskip}            
        $B_{s1}\rightarrow B_{s}\gamma$ & $70\pm10$ & $41\pm6$ & 0.02 & $220\pm31$ \\\noalign{\smallskip}
        $B_{s1}\rightarrow B^{*}_{s}\gamma$& $110\pm15$ & $0.20\pm0.02$ & 0.03 & $100\pm15$ \\\noalign{\smallskip}
        $B_{s1}\rightarrow B_{s0}^*\gamma$& $-$& $0.03\pm0.01$ & ? & ? \\\noalign{\smallskip}
  \hline
  \end{tabular}
  \label{tab:radiative}
  \end{table}

For the calculation of the hadronic decay widths, the formulation is almost exactly the same as that in Ref.~\cite{Cleven:2014oka} and is not repeated here (see Ref.~\cite{Cleven:2014oka} for details). There is, however, one crucial difference.

\begin{figure}[tb]
  \centering
  \includegraphics[width=8cm]{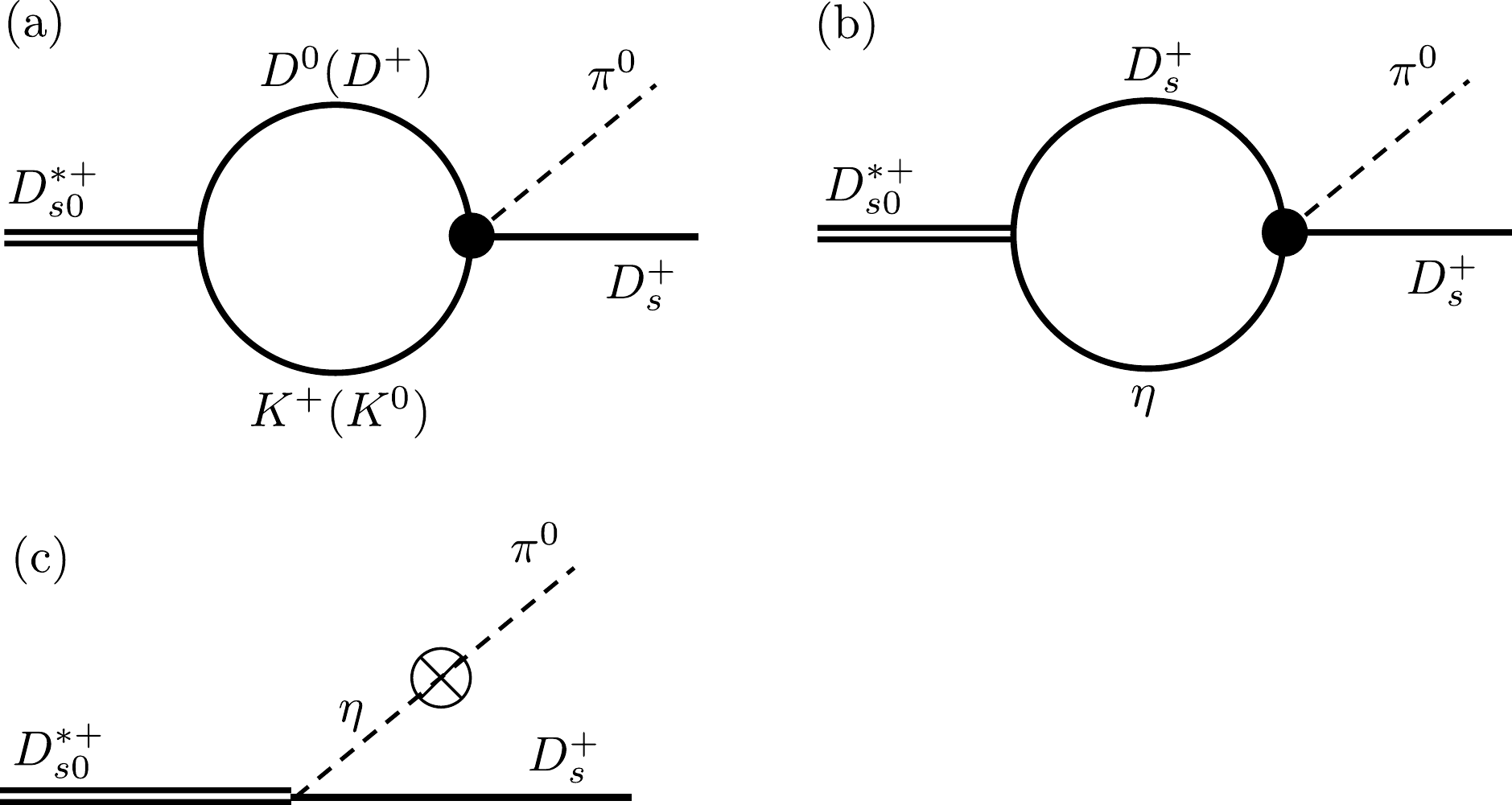}
  \caption{The diagrams for the hadronic decay mode $D_{s0}^{*}\to D_s^+\pi^0$. (a), with isospin conserving $D^{+(0)}K^{0(+)}\to D_s^+\pi^0$ four-point vertices, and (c) were considered in Ref.~\cite{Cleven:2014oka}.
  Here we consider all these three diagrams and four-point vertices with isospin breaking effects.
  }
  \label{fig:Diagram}
\end{figure}

\begin{table}[tb]
  \centering
  \caption{Masses and hadronic partial decay widths of $\dszero$, $\dsone$, $\bszero$ and $\bsone$ computed from searching for poles in the complex energy plane of the $T$-matrices in the particle basis. The mass and width correspond to the real part and twice the absolute values of the imaginary part of the pole location, respectively.}
  \begin{tabular}{lcc}
  \hline\noalign{\smallskip}    Meson & Mass [MeV] & Hadronic width (pole) [keV]         \\\noalign{\smallskip}
  \hline  
            $D_{s0}^{*}$ & $2318$ (fixed) &  $120^{+18}_{-\phantom{1}4}$ \\\noalign{\smallskip}
  $D_{s1}$ &$2458^{+15}_{-17}$ &$102^{+27}_{-11}$  \\\noalign{\smallskip}
  $B^{*}_{s0}$ &$5722\pm14$  &$75^{+24}_{-\phantom{2}9}$ \\\noalign{\smallskip}
  $B_{s1}$ & $5774\pm13$ & $74^{+23}_{-\phantom{2}8}$   \\\noalign{\smallskip}
  \hline
  \end{tabular}
  \label{tab:particle}
\end{table}

In Ref.~\cite{Cleven:2014oka}, the hadronic decay widths are split into two kinds of contributions, namely that from the $\pieta$ mixing, and  that from $DK$ loops due to the mass differences between the charged and neutral $D$ and $K$ mesons. They are shown in Fig.~\ref{fig:Diagram}~(c) and (a), respectively.
The hadronic width obtained in that way for the $D_{s0}^*(2317)$ is $(95\pm5)$~keV~\cite{Cleven:2014oka}.\footnote{The error quoted here is smaller than that in Ref.~\cite{Cleven:2014oka} as there was a typo in the code for the $\pieta$ mixing used therein.  }
The width may also be related to twice the imaginary part of the pole in the complex energy plane. For comparison, in Table~\ref{tab:particle} we also give the poles obtained by considering isospin breaking in the $T$-matrix in the particle basis (the channels are $D^+K^0, D^0K^+, D_s^+\eta$ and $D_s^+\pi^0$) using the same LECs.
One sees that the result for the $\dszero(2317)$ in Ref.~\cite{Cleven:2014oka} is smaller than that in Table~\ref{tab:particle}; the same also happens for the $\dsone(2460)$. 

We find that the main reason for the difference is that in the transition vertex $D^+K^0(D^0K^+)\to D_s\pi^0$ denoted by the solid circle in (a), there is also a subleading isospin breaking contribution, which was, however, neglected in Ref.~\cite{Cleven:2014oka}.
Here, we keep these isospin breaking terms in the four-point transition vertices ($D^+K^0(D^0K^+)\to D_s\pi^0$ and analogous transitions for other heavy-strange sectors) as derived from the next-to-leading order chiral Lagrangian~\cite{Guo:2008gp}, and consider further Fig.~\ref{fig:Diagram}~(b) for self-consistency.\footnote{It turns out that the contribution from Fig.~\ref{fig:Diagram}~(b) is negligible since the amplitude is proportional to the low-energy constant $h_0$, which is much smaller than the other $h_i$'s~\cite{Liu:2012zya}.} 
It is found that the contributions from loops increase sizeably compared with those given in Ref.~\cite{Cleven:2014oka}, and consequently the hadronic widths obtained in this way, see Table~\ref{tab:hadronicSU3f}, almost recover those from pole searching (Table~\ref{tab:particle}).

The results are consistent with all the available experimental information.
In Table~\ref{tab:comparison}, we show a comparison for the following ratios of partial widths to the available data and an update of the corresponding results in Ref.~\cite{Cleven:2014oka}:
\begin{equation}
\begin{split}
R_{1}&=\frac{\Gamma(D^{*}_{s0}\rightarrow D^{*}_{s}\gamma)}{\Gamma(D^{*}_{s0}\rightarrow D_{s}\pi^{0})},\quad R_{2}=\frac{\Gamma(D_{s1}\rightarrow D_{s}\gamma)}{\Gamma(D_{s1}\rightarrow D^{*}_{s}\pi^{0})},\\
R_{3}&=\frac{\Gamma(D_{s1}\rightarrow D^{*}_{s}\gamma)}{\Gamma(D_{s1}\rightarrow D^{*}_{s}\pi^{0})},\quad  R_{4}=\frac{\Gamma(D_{s1}\rightarrow D_{s0}^{*}\gamma)}{\Gamma(D_{s1}\rightarrow D^{*}_{s}\pi^{0})},\\
R_{5}&=\frac{\Gamma(D_{s1}\rightarrow D_{s}^{*}\pi^{0})}{\Gamma(D_{s1}\rightarrow D_{s}^{*}\pi^{0})+\Gamma(D_{s1}\rightarrow D^{*}_{s0}\gamma)},\\
R_{6}&=\frac{\Gamma(D_{s1}\rightarrow D_{s}\gamma)}{\Gamma(D_{s1}\rightarrow D_{s}^{*}\pi^{0})+\Gamma(D_{s1}\rightarrow D^{*}_{s0}\gamma)},\\
R_{7}&=\frac{\Gamma(D_{s1}\rightarrow D^{*}_{s}\gamma)}{\Gamma(D_{s1}\rightarrow D_{s}^{*}\pi^{0})+\Gamma(D_{s1}\rightarrow D^{*}_{s0}\gamma)},\\
R_{8}&=\frac{\Gamma(D_{s1}\rightarrow D_{s0}^{*}\gamma)}{\Gamma(D_{s1}\rightarrow D_{s}^{*}\pi^{0})+\Gamma(D_{s1}\rightarrow D^{*}_{s0}\gamma)} .\\
\end{split}
\label{eq:ratios}
\end{equation}
Our results are also consistent with the  the branching fraction of the $D_{s0}^*\to D_s\pi^0$, $(1.00^{+0.00}_{-0.20})$, measured recently by the BESIII Collaboration~\cite{Ablikim:2017rrr}.
\begin{table}[tb]
\centering
\caption{Comparison of the updated results for the  ratios of partial widths defined in Eq.~\eqref{eq:ratios} with the experimental data.  The central value of $R_2$ is fixed to reproduce the central measured value to determine the parameter for contact term in radiative decays (for details, see Ref.~\cite{Cleven:2014oka}).
}
\begin{tabular}{lcc}
\hline\noalign{\smallskip}  Ratio  &Our Result& Measured value \\\noalign{\smallskip}
\hline
      $R_{1}$ &$0.028\pm0.009$ &$<0.059$ \\\noalign{\smallskip}
       $R_{2}$ &$0.38 (\text{fixed})\pm0.08$ & $0.38\pm0.05$ \\\noalign{\smallskip}
      $R_{3}$ & $0.12\pm0.02$&$<0.16$ \\\noalign{\smallskip}
    $R_{4}$ &$0.028\pm0.006$ &$<0.22$  \\\noalign{\smallskip}
    $R_{5}$ & $0.97\pm0.01$&$0.93\pm0.09$ \\\noalign{\smallskip}
      $R_{6}$ &$0.37\pm0.08$ &$0.35\pm0.04$ \\\noalign{\smallskip}
     $R_{7}$ & $0.12\pm0.02$&$<0.24$ \\\noalign{\smallskip}
     $R_{8}$ & $0.027\pm0.004$& $<0.25$\\\noalign{\smallskip}
\hline
\end{tabular}
\label{tab:comparison}
\end{table}

To summarize, we update the results reported in Ref.~\cite{Cleven:2014oka} for the widths of the
lightest scalar heavy light mesons in the molecular
approach, since some input parameters are now known
with a higher accuracy, part of the isospin breaking sources was missing in Ref.~\cite{Cleven:2014oka} and, most importantly, the bottom-strange meson masses 
used therein were too small by about 100~MeV
compared to expectations deduced from HQFS. We traced this discrepancy to 
some HQFS violation in the regularization
procedure for the two meson loops coded in Ref.~\cite{Cleven:2014oka}.
The results presented here should be useful for the search of the lowest positive-parity bottom-strange mesons, with the $B_s\pi^0$ and $B_s\gamma$ being the preferred searching channels for $B_{s0}^*$ and $B_{s1}$, respectively, and revealing the internal structure of the charmed ones.

\section*{Acknowledgments}

This work is supported in part by the National Natural Science Foundation of China (NSFC) and the Deutsche For\-schungsgemeinschaft (DFG, German Research Foundation) through the funds provided to the Sino-German Collaborative Research Center ``Symmetries and the Emergence of Structure in QCD'' (NSFC Grant No. 12070131001, DFG Project-ID 196253076 -- TRR110), by NSFC under Grants No. 11835015, No.~12047503 and No.~11961141012, by the Chinese Academy of Sciences (CAS) under Grants No. XDPB15, No. XDB34030000 and No.~QYZDB-SSW-SYS013,
by CAS through a President's
International Fellowship Initiative (PIFI) (Grant No. 2018DM0034), by the VolkswagenStiftung
(Grant No. 93562), by the EU Horizon 2020 research and innovation programme, STRONG-2020 project
under grant agreement No 824093, and by the US Department of Energy under contract DE-SC0015393.



\begin{thebibliography}{99}
\bibitem{Cleven:2014oka}
M.~Cleven, H.~W.~Grie\ss{}hammer, F.-K.~Guo, C.~Hanhart and U.-G.~Mei\ss{}ner,
Eur. Phys. J. A \textbf{50} (2014) 149
[arXiv:1405.2242 [hep-ph]].

\bibitem{Aubert:2003fg}
B.~Aubert \textit{et al.} [BaBar],
Phys. Rev. Lett. \textbf{90} (2003) 242001
[arXiv:hep-ex/0304021 [hep-ex]].

\bibitem{Besson:2003cp}
D.~Besson \textit{et al.} [CLEO],
Phys. Rev. D \textbf{68} (2003) 032002
[erratum: Phys. Rev. D \textbf{75} (2007) 119908]
[arXiv:hep-ex/0305100 [hep-ex]].

\bibitem{Guo:2017jvc}
F.-K.~Guo, C.~Hanhart, U.-G.~Mei\ss{}ner, Q.~Wang, Q.~Zhao and B.-S.~Zou,
Rev. Mod. Phys. \textbf{90} (2018) 015004
[arXiv:1705.00141 [hep-ph]].

\bibitem{Nowak:2003ra}
M.~A.~Nowak, M.~Rho and I.~Zahed,
Acta Phys. Polon. B \textbf{35} (2004) 2377
[arXiv:hep-ph/0307102 [hep-ph]].

\bibitem{Bardeen:2003kt}
W.~A.~Bardeen, E.~J.~Eichten and C.~T.~Hill,
Phys. Rev. D \textbf{68} (2003) 054024
[arXiv:hep-ph/0305049 [hep-ph]].

\bibitem{Achasov:1979xc}
N.~N.~Achasov, S.~A.~Devyanin and G.~N.~Shestakov,
Phys. Lett. B \textbf{88} (1979) 367.

\bibitem{Hanhart:2007bd}
C.~Hanhart, B.~Kubis and J.~R.~Pelaez,
Phys. Rev. D \textbf{76} (2007) 074028
[arXiv:0707.0262 [hep-ph]].

\bibitem{Albaladejo:2016lbb}
M.~Albaladejo, P.~Fernandez-Soler, F.-K.~Guo and J.~Nieves,
Phys. Lett. B \textbf{767} (2017) 465
[arXiv:1610.06727 [hep-ph]].

\bibitem{Du:2017zvv}
M.-L.~Du, M.~Albaladejo, P.~Fern\'andez-Soler, F.-K.~Guo, C.~Hanhart, U.-G.~Mei\ss{}ner, J.~Nieves and D.~L.~Yao,
Phys. Rev. D \textbf{98} (2018) 094018
[arXiv:1712.07957 [hep-ph]].

\bibitem{Guo:2006fu}
F.-K.~Guo, P.-N.~Shen, H.-C.~Chiang, R.~G.~Ping and B.~S.~Zou,
Phys. Lett. B \textbf{641} (2006) 278
[arXiv:hep-ph/0603072 [hep-ph]].

\bibitem{Guo:2006rp}
F.-K.~Guo, P.-N.~Shen and H.-C.~Chiang,
Phys. Lett. B \textbf{647} (2007) 133
[arXiv:hep-ph/0610008 [hep-ph]].

\bibitem{Kolomeitsev:2003ac}
E.~E.~Kolomeitsev and M.~F.~M.~Lutz,
Phys. Lett. B \textbf{582} (2004) 39
[arXiv:hep-ph/0307133 [hep-ph]].

\bibitem{Lang:2015hza}
C.~B.~Lang, D.~Mohler, S.~Prelovsek and R.~M.~Woloshyn,
Phys. Lett. B \textbf{750} (2015) 17
[arXiv:1501.01646 [hep-lat]].

\bibitem{Faessler:2007gv}
A.~Faessler, T.~Gutsche, V.~E.~Lyubovitskij and Y.~L.~Ma,
Phys. Rev. D \textbf{76} (2007) 014005
[arXiv:0705.0254 [hep-ph]].

\bibitem{Gamermann:2007bm}
D.~Gamermann, L.~R.~Dai and E.~Oset,
Phys. Rev. C \textbf{76} (2007) 055205
[arXiv:0709.2339 [hep-ph]].

\bibitem{Faessler:2007us}
A.~Faessler, T.~Gutsche, V.~E.~Lyubovitskij and Y.~L.~Ma,
Phys. Rev. D \textbf{76} (2007) 114008
[arXiv:0709.3946 [hep-ph]].

\bibitem{Lutz:2007sk}
M.~F.~M.~Lutz and M.~Soyeur,
Nucl. Phys. A \textbf{813} (2008) 14
[arXiv:0710.1545 [hep-ph]].

\bibitem{Faessler:2008vc}
A.~Faessler, T.~Gutsche, V.~E.~Lyubovitskij and Y.-L.~Ma,
Phys. Rev. D \textbf{77} (2008) 114013
[arXiv:0801.2232 [hep-ph]].

\bibitem{Guo:2008gp}
F.-K.~Guo, C.~Hanhart, S.~Krewald and U.-G.~Mei{\ss}ner,
Phys. Lett. B \textbf{666} (2008) 251
[arXiv:0806.3374 [hep-ph]].

\bibitem{Xiao:2016hoa}
C.-J.~Xiao, D.~Y.~Chen and Y.-L.~Ma,
Phys. Rev. D \textbf{93} (2016) 094011
[arXiv:1601.06399 [hep-ph]].

\bibitem{Guo:2018kno}
X.-Y.~Guo, Y.~Heo and M.~F.~M.~Lutz,
Phys. Rev. D \textbf{98} (2018) 014510
[arXiv:1801.10122 [hep-lat]].

\bibitem{Fajfer:2015zma}
S.~Fajfer and A.~Prapotnik Brdnik,
Phys. Rev. D \textbf{92} (2015) 074047
[arXiv:1506.02716 [hep-ph]].

\bibitem{Fajfer:2016xkk}
S.~Fajfer and A.~Prapotnik Brdnik,
Eur. Phys. J. C \textbf{76} (2016) 537
[arXiv:1606.06943 [hep-ph]].

\bibitem{Liu:2012zya}
L.~Liu, K.~Orginos, F.-K.~Guo, C.~Hanhart and U.-G.~Mei{\ss}ner,
Phys. Rev. D \textbf{87} (2013)  014508
[arXiv:1208.4535 [hep-lat]].

\bibitem{Oller:2000fj}
J.~A.~Oller and U.-G.~Mei{\ss}ner,
Phys. Lett. B \textbf{500} (2001) 263
[arXiv:hep-ph/0011146 [hep-ph]].

\bibitem{Cleven:2010aw}
M.~Cleven, F.-K.~Guo, C.~Hanhart and U.-G.~Mei{\ss}ner,
Eur. Phys. J. A \textbf{47} (2011) 19
[arXiv:1009.3804 [hep-ph]].

\bibitem{Zyla:2020zbs}
P.~A.~Zyla \textit{et al.} [Particle Data Group],
PTEP \textbf{2020} (2020)  083C01 and the 2021 update.


\bibitem{Ablikim:2017rrr}
M.~Ablikim \textit{et al.} [BESIII],
Phys. Rev. D \textbf{97} (2018) 051103
[arXiv:1711.08293 [hep-ex]].

\end{thebibliography}

\end{document}